\documentclass[aps,pre,amsmath,amssymb,reprint]{revtex4-2}
\usepackage[utf8]{inputenc}
\usepackage{mfirstuc}
\usepackage{graphicx}
\usepackage{xcolor}
\usepackage[colorlinks=true, linkcolor=blue, citecolor=blue, urlcolor=blue]{hyperref}

\newcommand{\SK}[1]{\textcolor{black}{{#1}}}

\begin{document}
\title{The Universal Role of Fragility on the Yielding Transition of Active Glass under Oscillatory Shear}

\author{Arnab Mandal}
\affiliation{Tata Institute of Fundamental Research, 36/P, Gopanpally Village, Serilingampally Mandal, Ranga Reddy District, Hyderabad 500046, Telangana, India}

\author{Roni Chatterjee}
\affiliation{Tata Institute of Fundamental Research, 36/P, Gopanpally Village, Serilingampally Mandal, Ranga Reddy District, Hyderabad 500046, Telangana, India}

\author{Smarajit Karmakar}
\email{smarajit@tifrh.res.in}
\affiliation{Tata Institute of Fundamental Research, 36/P, Gopanpally Village, Serilingampally Mandal, Ranga Reddy District, Hyderabad 500046, Telangana, India}
\date{\today}

\begin{abstract}
\SK{The yielding transition marks the onset of macroscopic irreversible plastic deformation in amorphous solids and plays a central role in determining the mechanical stability, failure, and processing of a wide range of materials, including metallic glasses, colloidal suspensions, emulsions, and biological assemblies. Despite extensive research, the microscopic factors governing the nature of yielding, particularly the transition between brittle and ductile mechanical responses, remain poorly understood. Recent studies have identified kinetic fragility as a key parameter governing yielding in thermal/passive glasses. However, whether this connection persists in active glasses remains an open question. Here, using extensive molecular dynamics simulations of a Kob–Andersen glass former doped with Run-and-Tumble (RTP) active particles and subjected to oscillatory shear, we investigate how activity-induced changes in glassy dynamics influence yielding. We show that activity systematically reduces the kinetic fragility of the glass and consequently alters its mechanical response. The common yield point, $\gamma_c$, shared by poorly annealed samples decreases monotonically with activity and exhibits a power-law dependence on the high-temperature Arrhenius activation barrier. Increasing activity suppresses the dependence of the yield strain on thermal history and transforms the yielding response from brittle-like to increasingly ductile behaviour, characterized by smoother stress relaxation and reduced stress discontinuities. Moreover, the timescale to reach a steady state under oscillatory shear cycles near yielding continues to display critical-like power-law divergence, indicating that the nonequilibrium nature of activity does not alter the underlying critical character of the transition. Furthermore, active glasses develop broader, more diffuse shear bands, reflecting an enhanced spatial distribution of plastic deformation. Our results establish kinetic fragility as a unifying parameter governing yielding in both passive and active glasses and demonstrate that activity offers a powerful route to tune the mechanical response of amorphous materials.}
\end{abstract}

\maketitle

\noindent{\bf \large Introduction:} \SK{Understanding the mechanical response of amorphous solids is a problem of fundamental importance, both for basic science and for the wide range of structural and functional material -- metallic glasses, colloidal suspensions, dense emulsions -- that lack long-range crystalline order. In crystalline solids, plastic deformation is well understood in terms of the nucleation and motion of topological defects known as dislocations. The absence of such well-defined structural defects in amorphous systems makes it difficult to construct an analogous theoretical framework for their mechanical response \cite{FalkLanger_annrev2011, Fielding_Sollich_Cates2000, schuh2007, Bonn2017ModPhysRev}. The stress-strain behaviour of amorphous solids under both uniform and cyclic shear has been extensively studied \cite{Maloney_pre, maloney_prl2004, FalkLanger_pre, Leishangthem2017natcomm, ParmarPRX2019, priya2026prm}: at small deformations the response is elastic and reversible, but beyond a critical strain irreversible plastic rearrangements occur, mediated by localised structural events known as shear transformation zones (STZs) \cite{FalkLanger_pre, JS_Langer2008pre}. Despite sustained effort to understand the statistics and microscopic causes of these rearrangements \cite{Itamar_PKJaiswal_prl2016,pingua2025cascade, ParisiItamar_msingh_pnas2017, Smarajit_pre_2010, Smarajit2010RapidPRE}, a complete theoretical understanding of the yielding transition remains elusive.}

\SK{The nature of yielding is highly sensitive to the preparation history of the glass. Samples prepared from a high-temperature melt (poorly annealed) behave very differently from those prepared via extremely slow cooling rates (well-annealed), under both uniform and cyclic shear \cite{Ozawa_pnas2018, Bhaumik2021PNAS, BerthierOzawa_prl2020, Lamp2022JCP}. Under oscillatory shear, poorly annealed samples undergo mechanical annealing—progressively lowering their energy toward a threshold—with increasing strain amplitude, $\gamma_{max}$, before yielding. Well-annealed samples, by contrast, show no measurable pre-yield annealing and yield abruptly, via shear band formation. Above the yield point, the post-yield state is independent of preparation history (loss of memory \cite{Adhikari2018Memory, MemoryFormationSastry2019, Chatterjee_2026Memory}). A further remarkable feature of the yielding transition under cyclic shear is the divergence of the number of cycles, $\tau_{cy}$, required to reach steady state as the strain amplitude approaches the yield point, $\gamma_c$, from either side \cite{Bhaumik2021PNAS, MaityNatPhys2026}.}

\SK{It has recently been established, across a wide variety of model glass formers, that the kinetic fragility $K_\mathrm{VFT}$, which characterises how rapidly the relaxation time grows upon cooling, is the key parameter controlling the nature of the yielding transition under cyclic shear \cite{chatterjee2026role, Bhaumik2021PNAS, Roni_PRE}. Fragile glass formers, in which the relaxation time grows faster than Arrhenius, show strong sensitivity to the degree of annealing: the yield strain, $\gamma_y$, increases rapidly with decreasing parent temperature ($T_P$), with respect to the critical yield point, $\gamma_c$ and the stress drop, $\Delta \sigma$ at yielding becomes increasingly sharp, indicating a brittle-like failure of the sample. Strong glass formers, in which the relaxation time follows an Arrhenius temperature dependence, show minimal variation in yield strain during annealing and retain a ductile-like mechanical response. This fragility-controlled behaviour correlates very well with the growth of the energy barrier, $\Delta E$, with annealing: in fragile liquids, barriers grow more rapidly, driving a faster increase in the yield point and a larger stress overshoot. Whereas for strong glass-formers, the barrier grows more modestly within the studied annealing window, leading to very small changes in the yield strain. Whether this universality persists in the presence of active driving, which is itself known to alter fragility, remains an open question.}

\SK{Active matter systems are composed of self-driven entities that consume energy internally to generate motion, placing them in inherently out of equilibrium states and beyond the framework of standard equilibrium statistical mechanics \cite{Marchetti2013RMP}. Dense assemblies of active particles, including bacterial cytoplasm, dense cell layers, and synthetic self-propelled Janus colloids, exhibit slow glassy dynamics \cite{ThomasPNAS2011_glassycell, parry2014bacterialcytoplasm}, motivating the study of active glasses. A well-established minimal model for such systems introduces a fraction $\rho_a$ of Run-and-Tumble particles (RTPs) into a host glass former, where each active particle exerts a self-propulsion force of magnitude $f_0$ that reorients itself after a persistence time $\tau_p$ \cite{mandal2016active}. Within this framework, simulations have revealed a range of non-trivial dynamical signatures, including the rapid growth of dynamical heterogeneity \cite{kallol2023pnas, Subhodeep2022SoftMatter} and enhanced long-wavelength fluctuations \cite{Subhodeep2025natcomm}. Crucially, RTP activity has been shown to drive a systematic fragile-to-strong crossover in the glassy dynamics \cite{saroj_pnas_2018, mandal2016active}, reducing the kinetic fragility $K_\mathrm{VFT}$ and rendering the structural relaxation time increasingly Arrhenius with increasing activity.}

\SK{The mechanical response of active amorphous systems has begun to attract attention more recently. Studies have reported ductile and rounded yielding in active ultra-stable glasses \cite{Priya2026natcomm}, shear-induced ordering \cite{Rituparno2021pnas_actshear}, and activity-induced jamming \cite{ActiveJamming}, activity-induced brittle-ductile transition \cite{Sharma2025activity}. Despite these advances, the effect of activity-driven changes in fragility on the nature of the yielding transition -- in particular, the brittle-to-ductile crossover, the common yield point $\gamma_c$, and shear band morphology -- remains largely unexplored. Since fragility has recently been identified as the key parameter controlling yielding in passive glasses \cite{chatterjee2026role, Roni_PRE}, and since activity systematically tunes fragility, the active glass system offers a unique opportunity to probe this connection in an inherently non-equilibrium setting, which is the central aim of this article.}

\SK{In this paper, we address this gap directly. Using molecular dynamics simulations of the Kob-Andersen binary mixture with RTPs under oscillatory shear, we show that: (i) kinetic fragility controls the brittle-to-ductile transition in active glasses, extending the universality established for passive systems; (ii) activity provides direct and systematic control over the common yield point $\gamma_c$ through its effect on the Arrhenius activation energy $\Delta E$, with a novel power-law relation $\gamma_c \sim (\Delta E / k_B T_\text{on})^b$; and (iii) the more ductile stress response of active glasses is reflected in a broader and more spatially diffuse shear band. The rest of the paper is organised as follows. Section~\ref{sec:model} describes the model and simulation protocol. Section~\ref{sec:result} characterises the dynamical properties of the active glass. Sections \ref{sec:result}--\ref{sec:shearband} present our main results on yielding, the critical yield point, and shear band morphology. Finally, we conclude in Section~\ref{sec:conclusion}.}

\section{Models and Methods} \label{sec:model}
\noindent{\bf The Kob-Anderson Model:}
\SK{We simulate a binary mixture of Kob-Anderson liquid with $80:20$ composition \cite{KA_PhysRevE.51.4626}, where particles interact via a modified Lennard-Jones potential. 
\begin{equation*} \label{eqn2.2}
  \Phi_{\alpha \beta}(r) = 4\epsilon_{\alpha \beta} \left[ \left(\frac{\sigma_{\alpha \beta}}{r}\right)^{12} - \left(\frac{\sigma_{\alpha \beta}}{r}\right)^6  + C_0 + C_2 \left( \frac{r}{\sigma_{\alpha \beta}} \right)^2 \right],  
\end{equation*}
where $\alpha, \beta \in \{A, B\}$ denote two specific types of particles with interaction parameters $\epsilon_{AB}/\epsilon_{AA}=1.5, \epsilon_{BB}/\epsilon_{AA}=0.5$, and diameter ratios $\sigma_{AB}/\sigma_{AA}=0.80$ and $\sigma_{BB}/\sigma_{AA}=0.88$. These parameters are chosen to suppress crystallization at very low temperatures. The potential is truncated at $r_c=2.5\sigma_{\alpha \beta}$ and smoothened so that both $\Phi(r)$ and $\partial \Phi(r)/\partial r$ remain continuous, which uniquely determines the coefficients $C_0$ and $C_2$. We consider a system of $N=5000$ particles at a number density $\rho=1.2$ in a three-dimensional cubic simulation box. All results hereafter are reported in reduced units: length in units of $\sigma_{AA}$, energy in units of $\epsilon_{AA}$ and time in units of $\tau_{AA} = (m \sigma_{AA}^2/48 \epsilon_{AA})^{1/2}$.}  

\vskip +0.2in
\noindent{\bf Active Particle Dynamics:}
\SK{To introduce activity, we employed Run-and-Tumble particle (RTP) dynamics \cite{mandal2016active} in which activity is controlled by three parameters: active fraction ($\rho_a = N_{\text{act}}/N$), the self-propulsion force magnitude ($f_0$) and tumbling persistence time ($\tau_p$). The equation of motion for $i^{\text{th}}$ particle is
\begin{equation*}
    \begin{aligned}
        &\dot{\mathbf{r}}_i = \frac{\mathbf{p}_i}{m}, \\
        &\dot{\mathbf{p}}_i = - \mathbf{\nabla} \Phi_{\alpha \beta}(r) + n_i \mathbf{F}_i^{\text{act}}
    \end{aligned}
\end{equation*}
where $n_i$ is an indicator function that takes the value $0$ or $1$ depending on whether the particle is passive or active. The active force on the $i^{\text{th}}$ particle given by
\begin{equation*}
    \mathbf{F}_i^{\text{act}} = f_0 \left(k_i^x \hat{x} + k_i^y \hat{y} + k_i^z \hat{z} \right) 
\end{equation*}
where each active direction $k_i^\alpha$, $\alpha \in \{x,y,z\}$ are drawn from $\{+1,- 1\}$. The number of active particles is always chosen to be even, ensuring that the total centre-of-mass momentum vanishes component-wise, i.e. $\sum_i k_i ^{\alpha} = 0, \forall \alpha \in \{x,y,z\}$. Although $\tau_p$ does not appear explicitly in the equation of motion, each active particle exerts a constant force along its assigned $k_i^\alpha$ direction and tumbles randomly, selecting a new direction only after a time interval $\tau_p$.}

\SK{NVT simulations (constant number of particles, volume, and temperature) were carried out for three sets of activity parameters:  (i) $\rho_a = 0.20$ with $f_0 = 2.0$, (ii) $\rho_a = 0.10$ with $f_0 = 2.0$, and (iii) $\rho_a=0.10$ with $f_0 = 1.0$, along with a passive ($f_0=0$) reference case \cite{Roni_PRE}.}
The persistence time was fixed at $\tau_p = 1.0$ for all active cases except section~\ref{sec:diff_taup} where we present results for different 3 sets of $\tau_p=\{0.1, 1.0, 10.0\}$. 
\SK{The equations of motion were integrated with a time step of $dt=0.005$, and temperature was maintained using a three-chain Nos\'e–Hoover thermostat \cite{martyna1992nose}. For each activity set, the liquid was equilibrated at successively decreasing temperatures until the structural relaxation time $\tau_\alpha$ reached $10^6$ (in reduced units), defining the lowest accessible temperature for that set. At each state point, between $8$ and $12$ independent runs are performed for better averaging.}

\vskip +0.2in
\noindent{\bf Shear Protocol:}
\SK{After equilibrating the liquids at their respective parent temperatures ($T_P$), samples were quenched to $T_{\text{sim}} = 0.01$ using a fast cooling rate of $0.02$ per unit time. This rate is sufficiently fast that the structural memory of the parent temperature is retained. However, this rapid cooling introduces artefacts of ageing and annealing during subsequent shear flow, which are discussed in the later section. To impose shear flow along the $xy$ direction at temperature $T_{\text{sim}}$ and shear rate $\dot{\gamma}$, the Gaussian-thermostated SLLOD \cite{Todd_Daivis_2017, Pan_sllod} equations of motion were employed. For the $i^{\text{th}}$ particle,
\begin{equation*}
    \begin{aligned}
        &\dot{\mathbf{r}}_i = \frac{\mathbf{p}_i}{m_i} + \dot{\gamma}\, y_i\, \hat{x}, \\
        &\dot{\mathbf{p}}_i = \mathbf{F}_i - \dot{\gamma}\, p_i^y\, \hat{x} 
        - \alpha\, \mathbf{p}_i + n_i \mathbf{F}_i^{\text{act}},
    \end{aligned}
\end{equation*}
where the mass of each particle is set to unity, and $\alpha$ is the Gaussian thermostat multiplier that fixes the total kinetic energy at each time step. It is determined self-consistently as
\begin{equation*}
    \alpha = \frac{\displaystyle\sum_i \left[ 
    \left(\mathbf{F}_i + n_i \mathbf{F}_i^{\text{act}}\right) \cdot \mathbf{p}_i 
    - \dot{\gamma}\, p_i^x p_i^y \right]}{\displaystyle\sum_i |\mathbf{p}_i|^2}.
\end{equation*}
Cyclic shear with linearly increasing and decreasing strain amplitudes was applied to probe the system's mechanical response. The strain sequence
\begin{equation*}
    0 \;\xrightarrow{}\; \gamma_{\text{max}} \;\xrightarrow{}\; 0 
    \;\xrightarrow{}\; -\gamma_{\text{max}} \;\xrightarrow{}\; 0
\end{equation*}
constitutes one complete cycle. The shear rate was set to $\dot{\gamma} = 10^{-4}$. This value is sufficiently slow that most major stress drops are well resolved and successive plastic rearrangements remain temporally separated, ensuring that the system relaxes to a large extent between plastic events \cite{Smarajit_pre_2010}. Lees--Edwards boundary conditions were implemented to preserve periodicity in the shear-deformed simulation box.}

\section{Results} \label{sec:result}
\begin{figure*}[ht]
\centering
\includegraphics[width=1.0\linewidth]{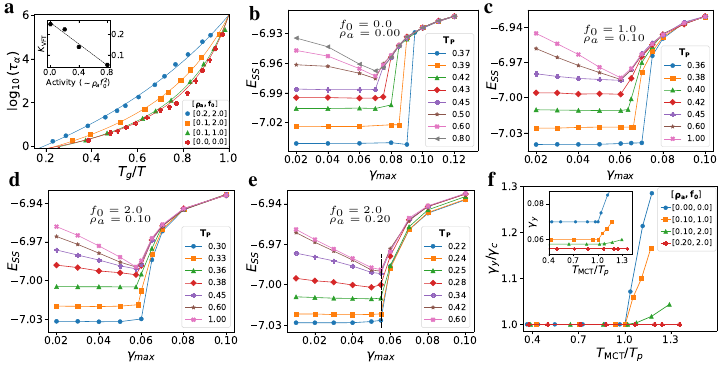}
\caption{\textbf{Effect of activity on the yielding phase portrait of an 
active glass.}
(a)~Angell plot of the structural relaxation time $\tau_\alpha$ as a function of rescaled inverse temperature $T_g/T$, where $T_g$ is the glass transition temperature defined by $\tau_\alpha(T_g) = 10^6$. Increasing the active fraction $\rho_a$ or the self-propulsion force $f_0$ systematically reduces the kinetic fragility $K_\mathrm{VFT}$ (inset), driving a fragile-to-strong crossover in the glassy dynamics.
(b)--(e)~Steady-state energy $E_{SS}$ as a function of strain amplitude $\gamma_\text{max}$ for the passive case (b) and three activity sets: $\rho_a=0.10$, $f_0=1.0$ (c); $\rho_a=0.10$, $f_0=2.0$ (d); and $\rho_a=0.20$, $f_0=2.0$ (e). In all panels, different curves correspond to different parent temperatures $T_P$. For the passive case (b), poorly annealed samples ($T_P > 0.435$) undergo mechanical annealing up to a common yield point $\gamma_c \approx 0.075$, while well-annealed samples ($T_P \leq 0.435$) exhibit no pre-yield annealing and yield discontinuously. With increasing dynamical annealing (decreasing $T_P$ below $0.435$), the yield point shifts progressively to larger strain amplitudes $\gamma_y > \gamma_c$. Panels (c)--(e) show that increasing activity reduces $\gamma_c$ and strongly suppresses the temperature-dependence of the yield point shift $\gamma_y/\gamma_c$; for the most active case (e), the yield point becomes independent of $T_P$ entirely. 
(f)~Relative yield point shift $\gamma_y/\gamma_c$ as a function of $T_\mathrm{MCT}/T_P$, comparing all activity sets. The progressive suppression of the yield point shift with increasing activity is consistent with the reduction in kinetic fragility, establishing fragility as the key parameter controlling the thermal dependence of yielding in active glasses.
Fig.~(b) adapted from \cite{Roni_PRE} with permission.
}

\label{fig:phase-portrait1}
\end{figure*}

\SK{After equilibrating the samples, the structural relaxation time $\tau_\alpha$ is computed from the overlap function $q(t)$ (defined in the appendix) at each parent temperature $T_P$. The temperature dependence of $\tau_\alpha$ is fitted to the Vogel--Fulcher--Tammann (VFT) equation, $\tau_\alpha = \tau_0 \exp\!\left(\frac{1}{K_\mathrm{VFT}\left(T/T_\mathrm{VFT} - 1\right)}\right),$
where $K_\mathrm{VFT}$ is the kinetic fragility and $T_\mathrm{VFT}$ is the temperature at which $\tau_\alpha$ formally diverges. The glass transition temperature $T_g$ is defined as the temperature at which $\tau_\alpha = 10^6$ (in reduced units), obtained by inverting the VFT fit. Fig.~\hyperref[fig:phase-portrait1]{\ref*{fig:phase-portrait1}(a)} shows the Angell plot \cite{AngellPlot1995} - $\tau_\alpha$ as a function of inverse temperature scaled by $T_g$ - for each activity set. Increasing activity, either by raising the active fraction $\rho_a$ or the self-propulsion force $f_0$, results in a systematic decrease in the kinetic fragility $K_\mathrm{VFT}$. As $K_\mathrm{VFT}$ decreases, the dynamics become increasingly Arrhenius, signalling a fragile-to-strong crossover driven by activity. To quantify the net activity across different parameter combinations, we adopt the mean-activity parameter $\Omega = \rho_a f_0^2\tau_p$, following Ref.~\cite{kallol2023pnas}. For the passive system, $K_\mathrm{VFT} = 0.252$, which undergoes a fivefold decrease to $K_\mathrm{VFT} = 0.052$ at the highest activity ($\rho_a = 0.20, f_0=2.0$). The inset demonstrates that $K_\mathrm{VFT}$ decreases nearly linearly with increasing activity $\Omega$.}

\begin{figure*} [ht]
    \centering
    \includegraphics[width=1.0\linewidth]{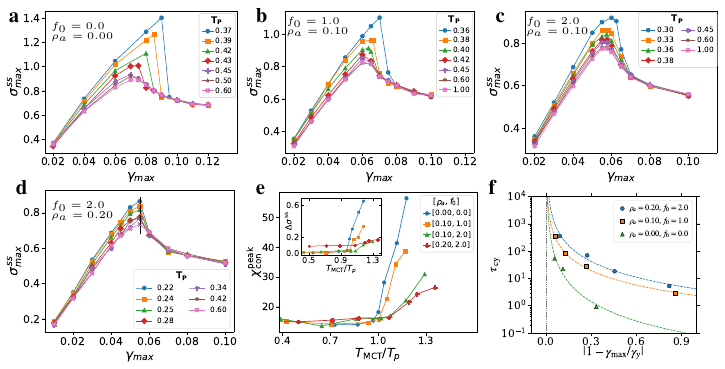}
    \caption{\textbf{Effect of activity on the brittle-to-ductile 
    transition.} 
    (a)--(d) Maximum steady-state stress $\langle\sigma\rangle_\text{max}^\text{ss}$ as a function of applied strain amplitude $\gamma_\text{max}$ for (a)~the passive case, (b)~$\rho_a = 0.10$, $f_0 = 1.0$, (c)~$\rho_a = 0.10$, $f_0 = 2.0$, and (d)~$\rho_a = 0.20$, $f_0 = 2.0$. For the passive glass, the stress drop at yielding becomes increasingly abrupt as $T_P$ decreases, characteristic of a brittle-like response. With increasing activity, the sharpness of the stress drop is progressively reduced even for the most well-annealed samples, indicating a systematic activity-driven shift toward more ductile behaviour.
    (e) Stress susceptibility $\chi_\text{con}^\text{peak} \equiv d\langle\sigma\rangle/d\gamma_\text{max}$ as a function of $T_\mathrm{MCT}/T_P$ for all activity sets. The peak value $\chi_\text{con}^\text{peak}$, which quantifies the abruptness of the stress drop, is strongly suppressed with increasing activity, confirming the brittle-to-ductile crossover. The inset shows the stress drop $\Delta\sigma$ -- computed as the difference between the last pre-yield and first post-yield stress values -- as a function of $T_\mathrm{MCT}/T_P$. The passive case shows a rapid increase with decreasing temperature, whereas the most active case remains nearly constant, consistent with the suppression of brittleness.
    (f) Cycle relaxation timescale $\tau_\text{cy}$, extracted by fitting a stretched exponential to the stroboscopic potential energy evolution (see Supplementary Sec.~S4 for details), as a function of $|1 - \gamma_\text{max}/\gamma_y|$ for well-annealed samples. Since no steady state is attained in the below-yield regime, $\tau_\text{cy}$ is presented for the above-yield regime only. The data are well described by the power-law form $\tau_\text{cy} = a\,|1 - \gamma_\text{max}/\gamma_y|^{b}$, demonstrating that the divergence of the relaxation timescale at the yield point persists in active glasses. 
    Fig.~(a) adapted from \cite{Roni_PRE} with permission.}
    \label{fig:stress_strain}
\end{figure*}

\SK{Each equilibrated sample was quenched to $T_\text{sim} = 0.01$, after which a large number of shear cycles were applied using the SLLOD algorithm. For poorly annealed samples, cyclic shear is known to induce mechanical annealing, progressively stabilising the glass \cite{Bhaumik2021PNAS, Leishangthem2017natcomm, Foffi_sastry_pre2013, Sastry2021PRL_EPM}. Sharma et al.\cite{Sharma2025activity} further demonstrated that activity acts analogously, introducing annealing into the system independently of external shear. When both cyclic shear and activity are present simultaneously, this raises an important question: can a steady state be achieved at all? In the below-yield regime, we find that the answer is no; the system continues to anneal slowly without reaching a steady state, indicating an activity and oscillatory shear-induced ageing. This behaviour is closely analogous to that observed under stress-controlled multidirectional cyclic shear \cite{Vishnu2023Multidirectional, goswami2025yielding}. In \cite{Sharma2025activity}, it was shown that activity introduces an effectively multidirectional perturbation that prevents the system from settling into a fixed limit cycle. The effect is most pronounced in poorly annealed glasses, which have a higher density of soft spots (shear transformation zones, or STZs)\cite{FalkLanger_pre, JS_Langer2008pre} that remain susceptible to both mechanical and active perturbations. To study this annealing process in a controlled and systematic manner, we adopt a uniform cutoff of $N_\text{cy} = 1000$ cycles for all temperatures $T > T_\mathrm{MCT}$, and all activity sets in the below-yield regime (details of this choice are discussed in the appendix). The average potential energy $E_\text{SS}$ as a function of strain amplitude $\gamma_\text{max}$, computed at this cutoff, is shown for a wide range of temperatures in Fig.~\ref{fig:phase-portrait1}.}

\SK{For the passive case Fig.~\ref{fig:phase-portrait1}(b), we recover the usual yielding phase portrait. Poorly annealed samples ($T > T_\mathrm{MCT}$) anneal progressively toward a threshold energy with increasing $\gamma_{max}$ before yielding at the critical yield strain, $\gamma_c$, while well-annealed samples show no measurable change in energy or structure prior to yielding discontinuously at $\gamma_y$. Once yielding occurs ($\gamma_\text{max} > \gamma_y$ or $\gamma_c$), the energy becomes independent of the degree of annealing - poorly and well-annealed samples collapse onto a single common value, signalling complete erasure of the thermal history. One notable feature of the passive case is that poorly annealed glasses share a common yield point $\gamma_c$, whereas for better dynamically annealed samples (decreasing $T_P$ below $T_\mathrm{MCT}$), the yield point shifts progressively to larger strain amplitudes $\gamma_y > \gamma_c$. Upon introducing activity, Figs.~\ref{fig:phase-portrait1}(c) and \ref{fig:phase-portrait1}(d) show results for the same active fraction $\rho_a = 0.10$ at two self-propulsion forces, $f_0 = 1.0$ and $f_0 = 2.0$, respectively. The most prominent change is a reduction in the common yield point $\gamma_c$ with increasing activity, which we attribute to the reduction in the Arrhenius activation energy barrier $\Delta E$; this connection is discussed in detail in the later section. A further notable feature is that the relative shift in the yield point, $\gamma_y/\gamma_c$, increases much more weakly with decreasing temperature than in the passive case. For $\rho_a = 0.10$, $f_0 = 1.0$, the common yield point is $\gamma_c = 0.06$, and even at the lowest accessible temperature ($T_\mathrm{MCT}/T = 1.17$), the yield point shifts to only $\gamma_y = 0.07$, a $16\%$ increase. By contrast, the passive system reaches a $30\%$ shift at a comparable reduced temperature ($T_\mathrm{MCT}/T = 1.175$). This suppression of the yield point shift is even more pronounced at higher activity: for $\rho_a = 0.10$, $f_0 = 2.0$, the shift remains as small as ${\sim}4\%$ even at $T_\mathrm{MCT}/T = 1.29$. These observations are consistent with recent findings \cite{chatterjee2026role, Roni_PRE} that identify kinetic fragility as the key parameter controlling the dependence of yielding on annealing in glassy systems. Finally, for the most active case ($\rho_a = 0.20$, $f_0 = 2.0$), the yield point is entirely independent of $T_P$, with $\gamma_y = \gamma_c$ for all degrees of dynamical annealing - a striking manifestation of the activity-driven fragile-to-strong crossover.}

\SK{To further characterise the change in yielding behaviour with activity, we examine the steady-state maximum stress $\sigma_\text{SS}^\text{max}$ as a function of strain amplitude $\gamma_\text{max}$ over a wide range of parent temperatures $T_P$. For the passive glass, decreasing $T_P$ leads to a progressively sharper, more discontinuous stress drop at yielding, a hallmark of a brittle-like mechanical response. As activity is increased, this stress drop diminishes progressively, and the stress-strain curve becomes more rounded, signalling a crossover toward ductile behaviour. For the most active case ($\rho_a = 0.20$, $f_0 = 2.0$), the response is fully ductile: the yield point $\gamma_y$ remains independent of $T_P$, consistent with the activity-driven fragile-to-strong crossover identified in the dynamics. This evolution from brittle to ductile behaviour is quantified by the stress susceptibility $\chi_\text{con} \equiv d\langle\sigma\rangle / d\gamma_\text{max}$, whose peak captures the abruptness of the stress drop -- a sharper, more discontinuous transition yields a larger, narrower peak. Fig.~\ref{fig:stress_strain}(e) shows the peak of $\chi_\text{con}$ as a function of $T_\mathrm{MCT}/T_P$ for all activity parameters studied in this work. For the passive glass, $\chi_\text{con}^\text{peak}$ grows most rapidly with decreasing temperature, while the most active case shows the slowest growth, with an almost flat dependence on $T_P$. This directly reflects the difference in kinetic fragility: a more fragile liquid develops a sharper yield transition upon annealing, whereas a stronger, more Arrhenius liquid remains relatively insensitive to thermal history. The inset of Fig.~\ref{fig:stress_strain}(e) corroborates this picture through the stress drop $\Delta\sigma$, defined as the difference between the last pre-yield and first post-yield stress values, plotted as a function of $T_\mathrm{MCT}/T_P$. The passive case shows a rapid increase as temperature decreases, whereas the most active case remains nearly constant --- confirming that kinetic fragility is the key parameter governing the brittle-to-ductile transition in active glasses.}

\vspace{0.2cm}
\noindent{\bf Divergence of the Cycle Relaxation Timescale}: \SK{We next characterise the cycle relaxation timescale $\tau_\text{cy}$, which measures the number of shear cycles required to reach steady state at a given strain amplitude $\gamma_\text{max}$. As discussed above, a steady state is attained only in the above-yield regime, since activity removes the absorbing state transition below the yielding transition; $\tau_\text{cy}$ is therefore extracted exclusively from above-yield data. The timescales are obtained from stretched exponential fits to the stroboscopic energy as a function of cycle number (Supplementary Sec.~S4). We present results for three activity sets: the passive reference, a low-activity case ($\rho_a = 0.10$, $f_0 = 1.0$), and the most active case ($\rho_a = 0.20$, $f_0 = 2.0$). For each set, $\tau_\text{cy}$ is fitted to the power-law form $\tau_\text{cy} \sim \left| 1 - \gamma_\text{max} / \gamma_y \right|^b$, demonstrating that $\tau_\text{cy}$ diverges as $\gamma_\text{max}$ approaches the yield point $\gamma_y$ from above, signalling a critical slowing down analogous to that observed in second order phase transitions. The fitted exponents are $b = -2.30 \pm 0.33$ (passive), $b = -1.93 \pm 0.17$ (most active, $\rho_a = 0.20$, $f_0 = 2.0$), and $b = -1.76 \pm 0.07$ (low activity, $\rho_a = 0.10$, $f_0 = 1.0$). These values are consistent with the exponent of $-2$ reported in Ref.~\cite{MaityNatPhys2026} for the passive case in the close vicinity of $\gamma_y$. The close agreement of the fitted exponents across all activity parameters demonstrates that the critical slowing down at the yield point is a robust feature of the yielding transition, preserved in the presence of activity. As a side note, one also sees that strong liquids typically take a larger number of oscillatory shear cycles to reach to the steady state - a fact that seems to be very generic across glass-formers both in the presence and absence of additional driving forces like activity.}

\section{Tuning Fragility via persistence time}\label{sec:diff_taup}
\begin{figure}[ht]
    \centering
    \includegraphics[width=1.0\linewidth]{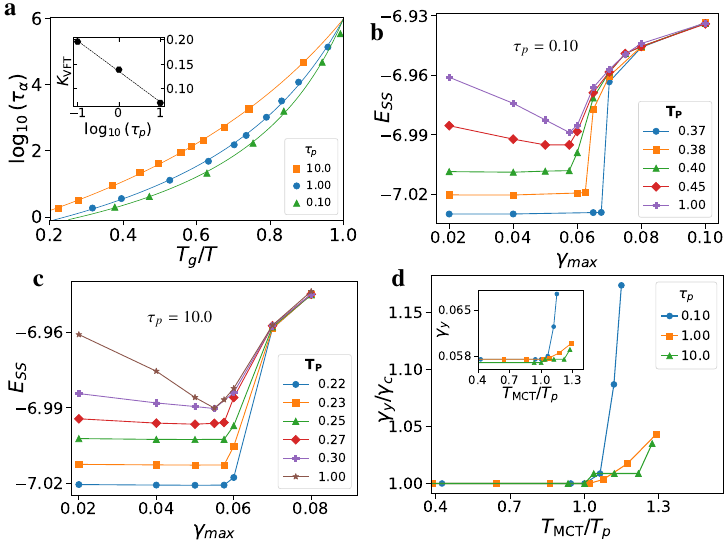}
    \caption{\textbf{Effect of persistent time $\mathbf{\tau_p}$ on yielding ($\mathbf{\rho_a=0.10, \ \ f_0=2.0}$).} 
    (a)~Angell plot of $\tau_\alpha$ vs $T_g/T$ for $\tau_p \in \{0.10, 1.0, 10.0\}$. Increasing $\tau_p$ systematically reduces $K_\mathrm{VFT}$, driving a fragile-to-strong crossover.
    (b),(c)~Steady-state energy $E_{SS}$ as a function of $\gamma_\text{max}$ for $\tau_p = 0.10$~(b) and $\tau_p=10.0$~(c); the $\tau_p=1.0$ result was already shown in Fig.~\ref{fig:phase-portrait1}(d). The temperature dependence of yielding is progressively suppressed with increasing $\tau_P$, and the transition becomes increasingly ductile.
    (d)~Relative yield point shift $\gamma_y/\gamma_c$ as a function of $T_\mathrm{MCT}/T_P$ for all three values of $\tau_p$. The suppression of the shift with increasing $\tau_p$ mirrors the reduction in $K_\mathrm{VFT}$, confirming that kinetic fragility controls the thermal dependence of yielding regardless of which activity parameter is varied.
    }
    \label{fig:diff_taup}
\end{figure}

\SK{For completeness, we showed that kinetic fragility can also be tuned by varying persistence time $\tau_p$ alone, independently of active fraction $\rho_a$ and self-propulsion force $f_0$. Fixing $\rho=0.10$ and $f_0=2.0$ we vary $\tau_p \in \{0.10, 1.0, 10.0\}$ and examine the fragility and the yielding behaviour under oscillatory shear. As shown in Fig.~\ref{fig:diff_taup}(a), variation of $\tau_p$ leads to a $\sim 2.5$-fold variation in $K_\mathrm{VFT}$, confirming that $\tau_p$ is also an important parameter in controlling the fragility of a glass-former. We note, however, that the sensitivity of fragility to $\tau_p$ depends on the overall activity level: for weaker active force $f_0=1.0$ with $\rho=0.10$, varying $\tau_p$ alone produces only a modest change in fragility. These behaviours have been reported, and an active random first-order transition (RFOT) theory \cite{saroj_pnas_2018} is found to rationalise these results.} 

\SK{The effect of fragility on the yielding transition seems to be universally the same irrespective of the origin of fragility across various glass formers. For $\tau_p=0.10$, the fragility is close to that of the passive case and the yield point shifts rapidly with increasing annealing. The results for $\tau_p=1.0$ were presented earlier in Fig.~\ref{fig:phase-portrait1}(d). For $\tau_p=10.0$, the fragility is substantially reduced, and the yield point becomes nearly independent of $T_P$. We further show the relative yield point shift $\gamma_y/\gamma_c$ as a function of $T_\mathrm{MCT}/T_P$ for all three values of $\tau_p$. For $\tau_p = 0.10$, the shift grows most rapidly as $T_P$ decreases below $T_\mathrm{MCT}$, while for $\tau_p=\{1.0, 10.0\}$, it remain nearly flat.}

\section{Dependence of the Common Yield Point on Activity} \label{sec:gamma_c}
\SK{We now examine what controls the critical yield point $\gamma_c$ shared by all poorly annealed samples. As already evident from the phase portrait, $\gamma_c$ decreases systematically with increasing activity. It has been established using the single-site elasto-plastic model (EPM) and simulation data \cite{chatterjee2026role} that $\gamma_c$ is controlled by the high-temperature Arrhenius activation energy $\Delta E$. To verify this connection in our active glass model, we determine $\gamma_c$ for a wide range of activity parameter combinations $(\rho_a, f_0)$, spanning from a passive reference to the most active case ($\rho_a = 0.30$, $f_0 = 2.0$), up to the point where the liquid begins to exhibit sub-Arrhenius dynamics. Activity systematically and monotonically reduces $\Delta E$, and consequently $\gamma_c$ decreases with increasing activity.}

\SK{For each activity parameter, $\gamma_c$ is plotted against the dimensionless activation barrier $\Delta E / k_B T_\text{on}$, where $k_B$ is the Boltzmann constant and $T_\text{on}$ is the onset temperature which demarcates the dynamical regimes that are influenced or dominated by the energy landscape. Typically, this temperature is computed from the temperature dependence of the inherent-structure energy or from the breakdown of the Stokes-Einstein relation \cite{Shiladitya2013SEB_JCP} in these systems. The activation energy $\Delta E$ is extracted by fitting the Arrhenius form,
\begin{equation*}
    \ln \tau_\alpha = \ln \tau_0 + \frac{\Delta E}{k_B T},
\end{equation*}
to $\tau_\alpha(T)$ restricted to the high-temperature regime (see Supplementary Materials Sec.~S5 for details). The onset temperature $T_\text{on}$ is determined from the inherent structure energy $\langle E_\text{IS} \rangle$ as a function of temperature. The inherent structure energy is known to follow $E_\text{IS} = E_\text{IS}^\infty + b/T$ at high temperatures, and $T_\text{on}$ is defined as the temperature below which $\langle E_\text{IS} \rangle$ begins to deviate from this form. We find that activity has a negligible effect on $T_\text{on}$ (Supplementary Materials Sec.~S5); accordingly, we adopt $T_\text{on} = 1.0$ for all activity sets.}

\begin{figure} [ht]
    \centering
    \includegraphics[width=0.85\linewidth]{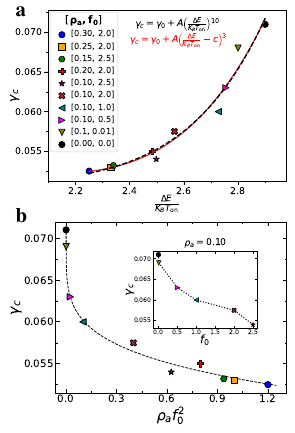}
    \caption{\textbf{Dependence of the critical yield point $\mathbf{\gamma_c}$ on activity.} All poorly annealed samples share a common yield point $\gamma_c$, which is controlled solely by the high-temperature Arrhenius activation energy barrier $\Delta E$. Increasing activity systematically reduces $\Delta E$, and consequently $\gamma_c$.
    (a) $\gamma_c$ computed for various combinations of $\rho_a$ and $f_0$, plotted against the dimensionless activation barrier $\Delta E / k_B T_\text{on}$, where $T_\text{on}$ is the onset temperature of Arrhenius behaviour. The data reveal a power-law dependence: even a modest level of activity substantially reduces $\gamma_c$, while beyond a threshold activity level, $\gamma_c$ saturates and becomes insensitive to further increases in activity. Black dashed curve is best-fit for $c=0.0, b \approx 10$, red continuous line is fit with $c=2.0, b \approx 3.0$
    (b) $\gamma_c$ as a function of the net activity parameter $\Omega = \rho_a f_0^2$, showing a nearly power law fall. The inset shows that for fixed $\rho_a = 0.10$, increasing $f_0$ monotonically reduces $\gamma_c$, confirming that both the active fraction and the self-propulsion force contribute independently to the suppression of the yield point.}
    \label{fig:gamma_c_change}
\end{figure}

\SK{Since $\gamma_c$ is highly sensitive to system size, preparation protocol, and shear rate, we hold all such parameters fixed and use $8$ independent realisations per activity set to determine $\gamma_c$ to a precision of $\pm 0.001$. The yield point is independently verified using both the zero-strain steady-state energy $E_\text{SS}$ and the cycle-to-cycle displacement $\Delta r$, following the protocol of Ref.~\cite{MaityNatPhys2026}.}

\SK{Having determined $\gamma_c$ for all activity sets, we fit the data as a function of $\Delta E / k_B T_\text{on}$ to the form
\begin{equation*}
    \gamma_c = \gamma_0 + A \left(\frac{\Delta E}{k_B T_\text{on}} - c\right)^b,
\end{equation*}
motivated by the observation of slow increment at small $\Delta E / k_B T_\text{on}$, followed by a sharp power-law increase at large values. The fits describe the data well, with an exponent $b \approx 10$ for $c = 0$, which reflects the extreme sensitivity of $\gamma_c$ to small changes in activation energy at low activity (large $\Delta E$). If we allow $c$ to vary, we find that $c = 2.0$ gives a similar fit with $b \simeq 3$. Although we note a strong correlation between the energy barrier and $\gamma_c$, in close agreement with the results reported in \cite{chatterjee2026role}, the functional dependence seems to be different. We don't yet understand the microscopic origin of this deviation for active amorphous solids.} 

\begin{figure*} [ht]
    \centering
    \includegraphics[width=1.0\linewidth]{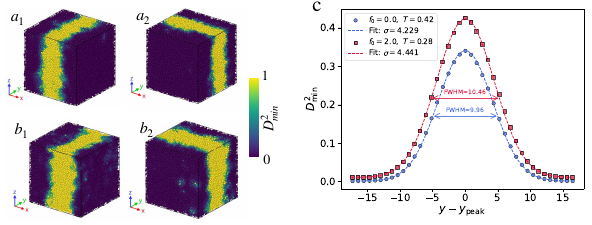}
    \caption{\textbf{Shear Band formation in active and passive glasses.}
     Non-affine displacement $D^2_\text{min}$, computed every $10$ cycles to improve shear band resolution, sheared in the $xy$-plane just above the yield point.
    (a$_1$, a$_2$) Passive glass prepared at $T_P = 0.42$, sheared at $\gamma_\text{max} = 0.075$ (just above $\gamma_y \sim 0.072$). Depending on the realisation, the shear band forms either along the shear direction ($x$) as in (a$_1$), or perpendicular to it ($y$) as in (a$_2$), illustrating the stochastic nature of shear band nucleation.
    (b$_1$, b$_2$) Active glass ($\rho_a = 0.20$, $f_0 = 2.0$) prepared at $T_P = 0.28$, sheared at $\gamma_\text{max} = 0.06$ (just above $\gamma_y \sim 0.055$), showing the same two orientations in different realisations. In both cases, the active glass exhibits a visibly broader shear band than its passive counterpart.
    (c) To quantify the band width, the $D^2_\text{min}$ (computed with 1 cycle gap) is averaged over both cycles and independent ensembles, and the simulation box is partitioned into slabs along the gradient direction to obtain a mean $D^2_\text{min}$ profile. Each profile is centred at its peak and fitted to a Gaussian,$A \exp\!\left(-y_\text{shifted}^2 / 2\sigma^2\right) + B$, to extract the band width $\sigma$. The active glass consistently exhibits a broader shear band compared to the passive case, as confirmed independently by the full width at half maximum (FWHM) of the fitted profiles.}
    \label{Fig :shearband}
\end{figure*}
\SK{To directly illustrate how activity tunes $\gamma_c$, we also plot $\gamma_c$ as a function of the net activity parameter $\Omega = \rho_a f_0^2\tau_p$, with $\tau_p = 1.0$. The yield point decreases monotonically with $\Omega$: an initial sharp decline at low activity gives way to a slow saturation at large $\Omega$, indicating that $\gamma_c$ becomes progressively less sensitive to further increases in activity. This saturation is consistent with the sub-Arrhenius crossover observed at high activity levels. $\gamma_c$ vs $\Omega$ can be fitted well by a power law form : $\gamma_c= a\Omega^b+\gamma_0$, where fitting gives $a=-0.175\pm0.0006$, $b=0.219\pm0.022$ and $\gamma_0=0.0707\pm0.0005$. For fixed $\rho_a = 0.10$, increasing $f_0$ produces a strictly monotonic reduction in $\gamma_c$, confirming that both $\rho_a$ and $f_0$ contribute independently to the suppression of the yield point through their combined effect on $\Delta E$. In this work, we restricted ourselves to persistence time, $\tau_p = 1.0$, to understand the regime of the active solids, which still falls in the linear response regime such that results can be contrasted to an effective equilibrium scenario with a suitable choice of effective temperature, although recent studies \cite{Priya2026natcomm}, suggest that rheological behaviour of active amorphous solids at small persistence time regime are not describable by an effective temperature like scenario. We expect that scenario with changing $\tau_p$ will be very interesting and important to study very systematically, as it is well-known that at large persistence time regime, these active systems show many interesting dynamical behaviour including intermittent dynamics, turbulence like behaviour etc.\cite{RituparnoNatCommExtremeActive}.}

\section{Shear Band Formation in Active Glass} \label{sec:shearband}

\SK{We conclude by examining how activity influences shear band formation. It is known that random pinning disorder which similarly decreases the fragility of a glass-forming liquid toward Arrhenius-like behaviour, completely suppresses shear band formation in passive glasses \cite{Roni_PRE, AnoopBhanu2025PRL}. Furthermore, network-like shear bands of increasing width have been reported in active ultra-stable glasses under simple shear \cite{priya2025pre, Priya2026natcomm}. Motivated by these observations and by the ductile stress response identified in the previous section, we investigate whether the active glass exhibits a spatially broader shear band compared to its passive counterpart.}

\SK{Additional simulations were performed with $N = 50{,}000$ particles at number density $\rho = 1.20$, giving a box length $L \approx 34.67\, \sigma_{AA}$, large enough to accommodate a well-defined shear band. Two sets of samples were prepared, each comprising $4$ independent realisations: a passive reference and a highly active system ($\rho_a = 0.20$, $f_0 = 2.0$), both equilibrated at parent temperatures just below their respective $T_\mathrm{MCT}$. Cyclic shear at $\gamma_\text{max} \gtrsim \gamma_y$ was applied until a steady state was reached. The non-affine displacement field $D^2_\text{min}$ \cite{FalkLanger_pre} was computed by accumulating particle displacements over $10$ successive cycles to enhance the spatial contrast of the shear band. Depending on the realisation, the band nucleates either along the shear direction or perpendicular to it, reflecting the stochastic nature of shear band nucleation along the principal axes of the shear deformation. To quantify the band width, the cycle-to-cycle $D^2_\text{min}$ profile was computed by partitioning the simulation box into slabs along the gradient direction. Each profile was recentred so that the slab of maximum cycle-averaged $D^2_\text{min}$ lies at the origin. The final $20$ cycles were averaged first, followed by the ensemble average; this ordering is essential to preserve the peak position at the centre. The resulting profile is well described by a Gaussian,
\begin{equation}
    D^2_\text{min}(y) = A \exp\!\left(-\frac{y_\text{shifted}^2}
    {2\sigma^2}\right) + B,
\end{equation}
from which the band width $\sigma$ and the full width at half maximum (FWHM) are extracted. The active glass exhibits a consistently broader shear band compared to the passive case, across all realisations. We attribute this to the more ductile mechanical response of the active glass: a less abrupt stress drop at yielding is associated with a more spatially extended plastic rearrangement zone, resulting in a broader and more diffuse band. This correlation between ductility and band width is physically intuitive -- a brittle system localises plastic flow into a sharp band, while a ductile system distributes it more homogeneously -- but its quantitative underpinning warrants more systematic investigation. In particular, larger system sizes are expected to make this distinction more pronounced, and a detailed finite-size analysis remains an interesting direction for future work.}

\section{Conclusion} \label{sec:conclusion}

\SK{We have investigated how the introduction of active noise in the form of Run-and-Tumble particles modifies the yielding behaviour of a model glass former under an oscillatory shear protocol, with particular focus on the role of kinetic fragility as a unifying parameter. Activity is known to drive a fragile-to-strong crossover in the glassy dynamics of the Kob-Andersen liquid; our central aim was to determine whether and how this dynamical change is reflected in the mechanical response under cyclic shear deformation.}

\SK{Our primary finding is that kinetic fragility $K_\mathrm{VFT}$ directly controls the nature of the yielding transition. For poorly annealed samples, the yielding behaviour is qualitatively similar across all activity levels -- all samples share a common yield point $\gamma_c$ and exhibit mechanical annealing below yield. As samples become better annealed, however, the fragility of the liquid determines how the yielding transition evolves: fragile liquids develop an increasingly brittle response with decreasing parent temperature $T_P$, characterised by a sharp stress drop and a rapidly growing stress susceptibility $\chi_\text{con}^\text{peak}$, whereas strong liquids -- rendered so by activity -- retain a ductile response largely independent of $T_P$. This behaviour is underpinned by the rate of energy barrier growth with annealing: in fragile liquids, barriers grow more rapidly, driving a faster increase in the yield point and a more pronounced stress overshoot. The universality of fragility as the controlling parameter of the brittle-to-ductile transition, recently established for passive glasses \cite{chatterjee2026role, Roni_PRE}, thus extends robustly to active amorphous solids.}

\SK{We additionally demonstrated that activity provides direct, systematic control over the common yield point $\gamma_c$ through its effect on the high-temperature Arrhenius activation energy $\Delta E$. Increasing activity -- whether through the active fraction $\rho_a$ or the self-propulsion force $f_0$ -- monotonically reduces $\Delta E$ and consequently $\gamma_c$. The dependence of $\gamma_c$ on the dimensionless barrier $\Delta E / k_B T_\text{on}$ follows a power law with an exponent $b \approx 3$, indicating extreme sensitivity of the yield point to small changes in activation energy at low activity levels, followed by saturation at large activity. This relationship can be expressed compactly in terms of the net activity parameter $\Omega = \rho_a f_0^2$, providing a practical single-parameter description of how activity tunes the mechanical threshold of the glass.}

\SK{We also found that randomly directed active forcing prevents the system from reaching an absorbing steady state in the below-yield regime, consistent with the known result that stochastically time-varying perturbations destroy limit cycles \cite{Sharma2025activity}. This effect is compounded by the analogy between RTP activity and multidirectional cyclic shear \cite{Vishnu2023Multidirectional}, both of which introduce continuous annealing that precludes a true periodic steady state.}

\SK{Finally, we showed that the ductile mechanical response of active glasses is directly reflected in the spatial structure of the shear band. Highly active glasses exhibit a consistently broader and more diffuse shear band compared to the passive case with comparable annealing history, as quantified by fitting the cycle-averaged $D^2_\text{min}$ profile to a Gaussian. We attribute this broadening to the less abrupt stress drop at yielding: a more ductile transition distributes plastic rearrangements more homogeneously across the system rather than localising them into a sharp band. While this correlation between ductility and band width is physically intuitive, its quantitative basis and in particular its dependence on system size and activity strength warrants more systematic investigation.}

\SK{Taken together, these results establish kinetic fragility as a universal parameter governing the mechanical response of active glasses: it controls the nature of the yielding transition, the sensitivity of the yield point to thermal history, and the spatial localisation of plastic flow. More broadly, our findings suggest that activity offers a practical, tunable handle for engineering the mechanical properties of amorphous solids --- shifting the yielding response from brittle to ductile and lowering the mechanical threshold without altering the underlying inter-particle interactions.}

\begin{acknowledgments}
A.M. thanks Rashmi Priya for useful discussion. We acknowledge the funding by intramural funds at TIFR Hyderabad from the Department of Atomic Energy (DAE) under Project Identification No. RTI 4007. SK would like to acknowledge the Swarna Jayanti Fellowship Grant Nos. DST/SJF/PSA01/2018-19 and SB/SFJ/2019-20/05 from the Science and Engineering Research Board (SERB) and Department of Science and Technology (DST). SK also acknowledges research support from MATRICES Grant MTR/2023/000079 from SERB.
\end{acknowledgments}


\bibliography{citation}

\end{document}


\title{Tuning the Yielding of Glass-Forming Liquids with Activity -- Supplementary Information}

\author{Arnab Mandal}
\email{arnabmandal@tifrh.res.in}
\affiliation{Tata Institute of Fundamental Research, 36/P, Gopanpally Village, Serilingampally Mandal, Ranga Reddy District, Hyderabad 500046, Telangana, India}

\author{Roni Chatterjee}
\affiliation{Tata Institute of Fundamental Research, 36/P, Gopanpally Village, Serilingampally Mandal, Ranga Reddy District, Hyderabad 500046, Telangana, India}

\author{Smarajit Karmakar}
\email{smarajit@tifrh.res.in}
\affiliation{Tata Institute of Fundamental Research, 36/P, Gopanpally Village, Serilingampally Mandal, Ranga Reddy District, Hyderabad 500046, Telangana, India}

\maketitle

\section{Dynamics of Active Glass}

To characterise the structural relaxation of the active glass, we compute two standard dynamical observables for the majority $A$-species: the mean squared displacement (MSD) and the self-overlap function $q(t)$. The MSD is defined as
\begin{equation*}
    \left\langle \Delta r^2(t) \right\rangle = \frac{1}{N_A} 
    \sum_{i=1}^{N_A} \left| \mathbf{r}_i(t + t_0) - 
    \mathbf{r}_i(t_0) \right|^2,
\end{equation*}
where $\mathbf{r}_i(t)$ is the position of particle $i$ at time $t$, $t_0$ is the reference time origin, and the average is taken over multiple time origins to improve statistical accuracy.

The self-overlap function $q(t)$ is defined as
\begin{equation*}
    q(t) = \frac{1}{N_A} \sum_{i=1}^{N_A} 
    \Theta\!\left(a - \left|\mathbf{r}_i(t + t_0) - 
    \mathbf{r}_i(t_0)\right|\right),
\end{equation*}
where $\Theta(x) = 1$ if $x > 0$ and $\Theta(x) = 0$ otherwise, so that $q(t)$ counts the fraction of particles that have not moved beyond a distance $a$ from their initial positions. The cutoff $a=0.3\sigma_\text{AA}$ is chosen such that short-time vibrational motion within the cage does not contribute to $q(t)$, ensuring that only genuine structural relaxation is captured. The structural relaxation time $\tau_\alpha$ is extracted from $q(t)$ as the time at which $q(\tau_\alpha) = 1/e$. 

\begin{figure}[ht]
    \centering
    \includegraphics[width=1.0\linewidth]{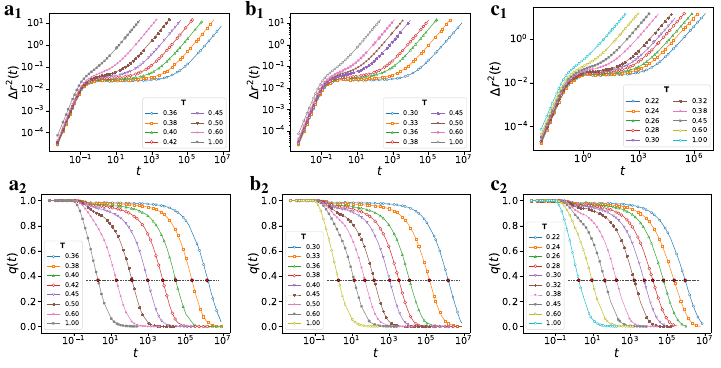}
    {\linespread{0.9}\selectfont
    \caption{Panels $(\text{a}_1,\text{a}_2)$, $(b_1,b_2)$, and $(c_1,c_2)$ show the mean squared displacement $\langle \Delta r^2(t) \rangle$ and self-overlap function $q(t)$, respectively, for $\rho_a = 0.10, f_0 = 1.0;$ $\rho_a = 0.10, f_0 = 2.0$; and $\rho_a = 0.20, f_0 = 2.0$}
    
    \label{fig:msd_qoft_all}}
\end{figure}

\section{Relaxation Data}
\begin{figure}[ht]
    \centering
    \includegraphics[width=0.80\linewidth]{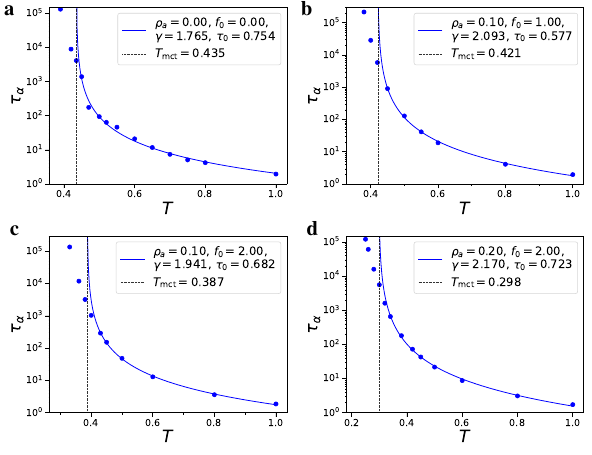}
    {\linespread{0.9}\selectfont
    \caption{Power-law fits of the structural relaxation time $\tau_\alpha(T)$ to the MCT form $\tau_\alpha = \tau_0(T - T_\mathrm{MCT})^{-\gamma}$ for all activity sets: (a) passive, (b) $\rho_a = 0.10$, $f_0 = 1.0$, (c) $\rho_a = 0.10$, $f_0 = 2.0$, and (d) $\rho_a = 0.20$, $f_0 = 2.0$. Solid lines denote the fits, restricted to the high-temperature regime. The extracted $T_\mathrm{MCT}$ values are used throughout the main text as the reference temperature separating poorly and 
    well-annealed samples.}
    \label{fig:mct_fit_all}}
\end{figure}

Within mode-coupling theory (MCT), the structural relaxation time $\tau_\alpha$ diverges as a power law upon approaching the MCT temperature $T_\mathrm{MCT}$ from above, $ \tau_\alpha = \tau_0 \left( T - T_\mathrm{MCT} \right)^{-\gamma},$
where $\tau_0$ and $\gamma$ are fitting parameters. The temperature $T_\mathrm{MCT}$ serves as a natural reference point in our analysis, as the shear response of the glass changes qualitatively in its vicinity (discussed in the main text). For each activity set, $T_\mathrm{MCT}$ is determined by fitting the above power-law form to $\tau_\alpha(T)$ in the high-temperature regime, where MCT is expected to hold. The resulting fits are shown in Fig.~\ref{fig:mct_fit_all}.

\section{Contribution of Active force to Stress}

In the SLLOD framework, the instantaneous $xy$-component of the stress tensor receives contributions from three distinct terms:
\begin{equation*}
\sigma_{xy}=\frac{1}{V}
\left(
\underbrace{\sum_i v_x^i v_y^i}_{\text{kinetic}} +
\underbrace{\sum_{i,j} f_{ij}^x\, y_{ij}}_{\text{virial}} +
\underbrace{\sum_i n_i\, y_i\, f_0^{\text{act},x}}_{\text{active}}
\right)
\end{equation*}

\begin{figure}[hb]
    \centering
    
    \includegraphics[width=1.0\linewidth]{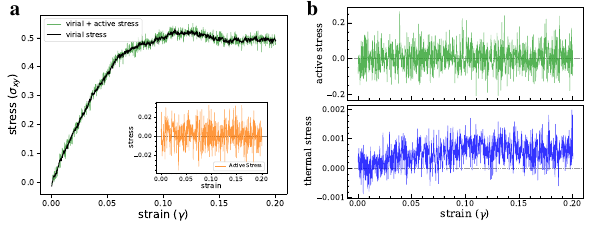}
    {\linespread{0.9}\selectfont
    \caption{Decomposition of the stress tensor $\sigma_{xy}$ into its constituent contributions for the highest activity case 
    ($\rho_a = 0.20$, $f_0 = 2.0$) at parent temperature $T_P = 1.0$.
    (a) Stress--strain curves under uniform shear comparing the full stress (virial $+$ active, green) and the virial stress alone (black), averaged over $32$ independent realisations. The two curves are indistinguishable on this scale, demonstrating that the active contribution is negligible after ensemble averaging. The inset shows the active stress contribution in isolation; it fluctuates about zero (dashed line) with magnitude $\mathcal{O}(0.02)$, suppressed from the single-realisation value by the expected $1/\sqrt{N_\text{real}}$ factor.
    (b) Active stress (top) and thermal (kinetic) stress (bottom) as functions of strain for a single realisation. The active stress fluctuates about zero with amplitude $\mathcal{O}(0.2)$, while the thermal stress is negligible at $\mathcal{O}(10^{-3})$ even for a single realisation. Note the difference in $y$-axis scales between the two subpanels. Together, panels (A) and (B) confirm that $\sigma_{xy}$ is well approximated by the interparticle viral term alone, and all stress values reported hereafter refer to this contribution.}
    \label{fig:stress_contri}}
\end{figure}

where $V$ is the system volume, $v_\alpha^i, \ \ (\alpha \in \{x,y,z\})$ are the velocity components of particle $i$, $f_{ij}^x$ is the x component of the interparticle force between particles $i$ and $j$, and $y_{ij}$ denotes the $y$-component of their separation vector. $n_i \in \{0,1\}$ is the activity indicator, $f_0^{\text{act},x}$ is self-propelled active force as defined before. We now argue that both the kinetic and active contributions are negligible compared to the virial term under our simulation conditions. Since all production runs are performed at $T_\text{sim} = 0.01$, the particle velocities are very small, and the kinetic contribution -- upon normalisation by the system volume -- gives a negligible thermal stress, as verified in Fig.~\ref{fig:stress_contri}. The active contribution vanishes for a more structural reason. Active particles are distributed approximately homogeneously throughout the simulation box, and their propulsion directions are constrained so that the total centre-of-mass momentum is conserved, i.e. $\sum_i k_i^\alpha = 0$ for all $\alpha$. As a result, the active stress $\sigma_{xy}^\text{act}$ fluctuates about zero rather than accumulating a systematic bias, shown in Fig.~\ref{fig:stress_contri}. For a single realisation, the magnitude of this fluctuation is of order $\mathcal{O}(0.2)$; however, upon averaging over $32$ independent realisations, it reduces to $\mathcal{O}(0.02)$, consistent with the expected $1/\sqrt{N_\text{real}}$ suppression of uncorrelated fluctuations. We therefore conclude that $\sigma_{xy}$ is well approximated by the interparticle contribution alone, and all stress values reported in the main text refer to this virial term.

\section{Stroboscopic Energy Evolution and Steady-State Analysis} \label{sec:steady-state-energy-analysis}

In this section we characterise the evolution of the potential energy sampled stroboscopically — at the end of each shear cycle — as a function of the number of cycles $N_\text{cy}$. For clarity, we present only the two extreme cases: the highest $T_P$ (most poorly annealed) and the lowest $T_P$ (most well-annealed) sample for each activity set, along with the passive reference. Since the finite-temperature SLLOD equations of motion are employed, thermal fluctuations introduce a persistent ageing effect throughout the simulation. Activity compounds this further, delaying the onset of a steady state by introducing additional annealing into the system \cite{sharma2025activity}. Consequently, for poorly annealed samples in the below-yield regime, no steady state is reached even after a large number of cycles, regardless of activity strength. To enable a consistent comparison across all activity sets and temperatures, we fit the stroboscopic energy $E(\gamma_\text{max}=0, N_\text{cy})$ as a function of cycle number to a stretched exponential,
\begin{equation*}
    E(N_\text{cy}) = E_0 + b \exp{\left[- \left( 
    \frac{N_\text{cy}}{\tau_\text{cy}}\right)^\beta \right]},
\end{equation*}
where $E_0$ is the extrapolated steady-state energy, $b$ is the amplitude, $\tau_\text{cy}$ is the cycle relaxation timescale, and $\beta$ is the stretching exponent. For poorly annealed samples in the below-yield regime, where no steady state is attained, we instead collect energies for the phase portrait at a uniform cutoff of $N_\text{cy} = 1000$ cycles (see main text). In the above-yield regime, the behaviour is strikingly different: after an initial transient, the energy of poorly annealed samples fluctuates about a well-defined mean value, which is used to construct the phase portrait.

The cycle relaxation timescale $\tau_\text{cy}$, extracted from the stretched exponential fit, is meaningful only in the above-yield regime, where a steady state is achieved. In the below-yield regime, the fit does not converge to a physical plateau and $\tau_\text{cy}$ carries no dynamical significance. For well-annealed samples, the behaviour mirrors that observed under AQS-driven shear \cite{Bhaumik2021PNAS}: no mechanical annealing occurs below the yield point, while above it, the energy increases abruptly and reaches a steady state.

\begin{figure} [ht]
    \centering
    \includegraphics[width=1.0\linewidth]{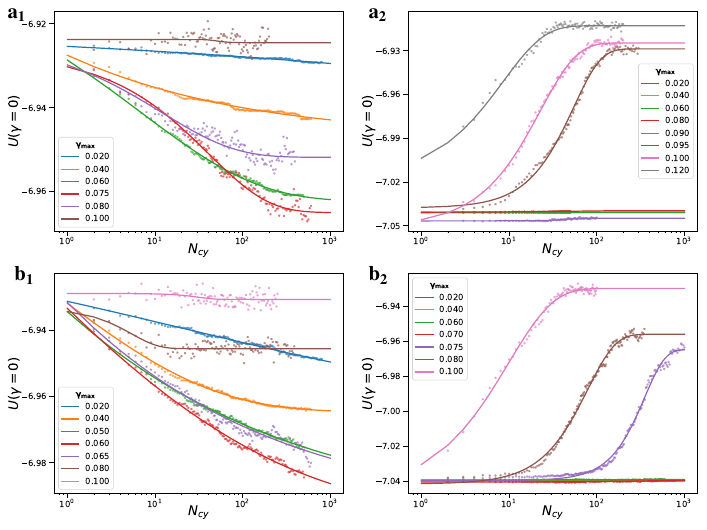}
    {\linespread{0.9}\selectfont
    \caption{Stroboscopic potential energy $E(N_\text{cy})$ sampled at the end of each shear cycle for varying strain amplitude $\gamma_\text{max}$, for the passive case. ($\text{a}_1$)~Poorly annealed sample ($T_P = 1.0$) and ($\text{a}_2$)~well-annealed sample ($T_P = 0.37$) (Fig. credit~\cite{Roni_PRE}), averaged over $12$ independent ensembles. ($\text{b}_1$)~Poorly annealed sample ($T_P = 1.0$) and ($\text{b}_2$)~well-annealed sample ($T_P = 0.36$) for $\rho_a = 0.10$, $f_0 = 1.0$. In the below-yield  regime, poorly annealed samples show no steady state, while well-annealed samples fluctuate about a well-defined mean above yield.}
    \label{fig:zero_strain_1}}
\end{figure}

\begin{figure} [ht]
    \centering
    \includegraphics[width=1.0\linewidth]{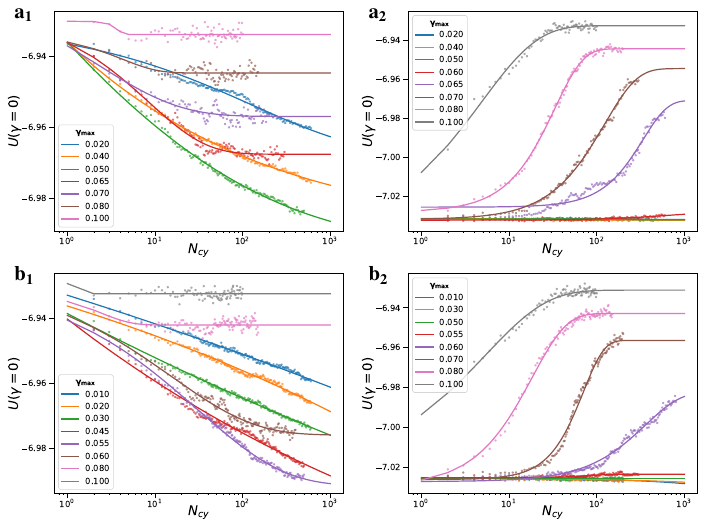}
    {\linespread{0.9}\selectfont
    \caption{Stroboscopic potential energy $E(N_\text{cy})$ as a function of cycle number for varying strain amplitude $\gamma_\text{max}$, for higher activity sets. ($\text{a}_1$)~Poorly annealed sample ($T_P = 1.0$) and ($\text{a}_2$)~well-annealed sample ($T_P = 0.37$) for $\rho_a = 0.10$, $f_0 = 2.0$. ($\text{b}_1$)~Poorly annealed sample ($T_P = 1.0$) and ($\text{b}_2$)~well-annealed sample ($T_P = 0.36$) for $\rho_a = 0.20$, $f_0 = 2.0$. The qualitative behaviour is consistent across activity sets; increasing activity further delays the approach to steady state in the below-yield regime.}
    \label{fig:zero_strain_2}}
\end{figure}

\section{Activation energy barrier and onset temperature}
The high-temperature activation energy barrier $\Delta E$ is determined by fitting the structural relaxation time $\tau_\alpha$ as a function of inverse temperature $1/T$ to the Arrhenius form, $\ln \tau_\alpha = \ln \tau_0 + \frac{\Delta E}{k_B T}$, where the slope of the linear fit in the $\ln\tau_\alpha$ vs $1/T$ representation gives $\Delta E / k_B$, and the fitting is restricted to the high-temperature regime where the Arrhenius form holds.

\begin{figure} [ht]
    \centering
    \includegraphics[width=\linewidth]{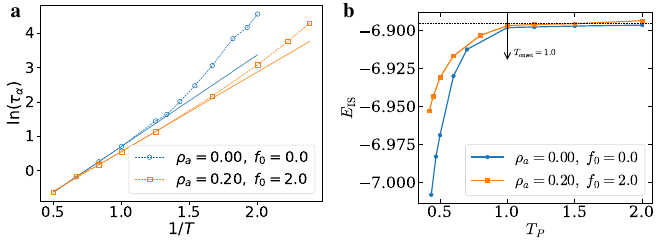}
    {\linespread{0.9}\selectfont
    \caption{Determination of the Arrhenius activation energy $\Delta E$ and onset temperature $T_\text{on}$.
    (a) Structural relaxation time $\ln\tau_\alpha$ as a function of inverse temperature $1/T$ for the passive case and the most active case ($\rho_a = 0.20$, $f_0 = 2.0$). Solid lines are Arrhenius fits restricted to the high-temperature regime; the slope gives $\Delta E / k_B$. Activity systematically reduces the activation barrier: $\Delta E = 2.90$ for the passive case, decreasing to  $\Delta E = 2.484$ for $\rho_a = 0.20$, $f_0 = 2.0$.
    (b) Inherent structure energy $\langle E_\text{IS} \rangle$ as a function of temperature, shown for both cases. At high temperatures, $\langle E_\text{IS} \rangle$ follows a constant plateau; below a characteristic temperature it begins to decrease, marking the onset of landscape-influenced dynamics. Both cases yield $T_\text{on} \approx 1.0$ (in reduced units), confirming that activity has negligible effect on the onset temperature.}
    \label{fig:arrhenius_gamma_c}}
\end{figure}

The onset temperature $T_\text{on}$ is determined from the temperature dependence of the inherent structure energy $\langle E_\text{IS} \rangle$. At high temperatures, $\langle E_\text{IS} \rangle$ is approximately constant, following the form $E_\text{IS} = E_\text{IS}^\infty + b/T$. Below a characteristic temperature, it begins to decrease as the liquid falls out of equilibrium and explores progressively deeper regions of the energy landscape. $T_\text{on}$ is defined as the temperature below which $\langle E_\text{IS} \rangle$ begins to deviate from its high-temperature plateau value $E_\text{IS}^\infty$. We find that activity has no significant effect on $T_\text{on}$. As shown in Fig.~\ref{fig:arrhenius_gamma_c}, the high-temperature $\langle E_\text{IS} \rangle$ plateau is nearly identical for the passive case and the most active case ($\rho_a = 0.20$, $f_0 = 2.0$), and both yield $T_\text{on} \approx 1.0$ (in reduced units). We therefore adopt $T_\text{on} = 1.0$ as a common value for all activity sets in the computation of the dimensionless activation barrier $\Delta E / k_B T_\text{on}$.

\bibliography{citation}